\documentclass[sn-basic]{sn-jnl}

\usepackage{graphicx}%
\usepackage{multirow}%
\usepackage{amsmath,amssymb,amsfonts}%
\usepackage{amsthm}%
\usepackage{mathrsfs}%
\usepackage[title]{appendix}%
\usepackage{xcolor}%
\usepackage{textcomp}%
\usepackage{manyfoot}%
\usepackage{booktabs}%
\usepackage{algorithm}%
\usepackage{algorithmicx}%
\usepackage{algpseudocode}%
\usepackage{listings}%

\usepackage{bm}
\usepackage{adjustbox}
\usepackage{natbib}

\DeclareMathOperator*{\argmax}{\arg\,max}

\begin{document}

\title[Deep Reinforcement Learning with Positional Context for Intraday Trading]{Deep Reinforcement Learning with Positional Context for Intraday Trading}


\author*[1]{\fnm{Sven} \sur{Goluža}}\email{sven.goluza@fer.hr}

\author[1]{\fnm{Tomislav} \sur{Kovačević}}\email{tomislav.kovacevic@fer.hr}

\author[1]{\fnm{Tessa} \sur{Bauman}}\email{tessa.bauman@fer.hr}

\author[1]{\fnm{Zvonko} \sur{Kostanjčar}}\email{zvonko.kostanjcar@fer.hr}

\affil[1]{\orgname{University of Zagreb, Faculty of Electrical Engineering and Computing}, \orgdiv{Laboratory for Financial and Risk Analytics (\href{lafra.fer.hr}{lafra.fer.hr}}),
\orgaddress{\street{Unska 3}, \postcode{10000} \city{Zagreb}, \country{Croatia}}}


\abstract{Deep reinforcement learning (DRL) is a well-suited approach to financial decision-making, where an agent makes decisions based on its trading strategy developed from market observations. Existing DRL intraday trading strategies mainly use price-based features to construct the state space. They neglect the contextual information related to the position of the strategy, which is an important aspect given the sequential nature of intraday trading. In this study, we propose a novel DRL model for intraday trading that introduces positional features encapsulating the contextual information into its sparse state space. The model is evaluated over an extended period of almost a decade and across various assets including commodities and foreign exchange securities, taking transaction costs into account. The results show a notable performance in terms of profitability and risk-adjusted metrics. The feature importance results show that each feature incorporating contextual information contributes to the overall performance of the model. Additionally, through an exploration of the agent's intraday trading activity, we unveil patterns that substantiate the effectiveness of our proposed model.}

\keywords{Deep reinforcement learning, Machine learning, Quantitative finance, Financial decision-making, Intraday trading}



\maketitle

\section{Introduction}\label{sec:intro}
With the advances in computational power and expanded datasets, there has been an increasing trend toward data-driven approaches to financial decision-making. Reinforcement learning (RL) is a popular data-driven approach to learning sequential decision-making, where agents learn through interaction with the environment by receiving feedback for their actions and improving their decision-making using their experience \citep{Sutton2018}. Deep reinforcement learning (DRL) combines the representational ability of deep learning with the trial-and-error learning of RL, and such an approach is naturally well-suited for financial decision-making problems. One of them is the development of intraday trading strategies, specifically designed to execute trades within a single day and mitigate overnight risk by refraining from holding assets overnight. These strategies enhance market efficiency and liquidity by actively contributing to real-time price formation and increasing overall trading activity \citep{Hendershott2011}.

Trading strategies are predetermined mathematical rules derived from market observations used to place trades, with the aim of generating profit while minimizing risk \citep{Narang2013}. The components of a trading strategy can vary depending on its particular implementation, but generally include essential elements such as a forecast module and an allocation module.
The forecast module is designed to predict future asset movements by analyzing historical data. This can be formulated as either a regression task, with the dependent variable being the magnitude (e.g., asset return), or as a classification task, with the dependent variable representing the direction (e.g., trend class).
The allocation module maps these forecasts into actual trades. However, this process poses a challenge due to the typically shorter time horizon of forecasts compared to the persistence of price trends \citep{Lim2019}. In existing literature, this issue is often addressed with a simplistic mapping, and such an approach is frequently insufficient, highlighting the necessity for a more robust allocation module. This module is crucial in establishing an effective mapping and in making trading decisions based on forecasts, risk considerations, and potential transaction costs \citep{Lim2019}.  Without this module, the direct trading approach following a supervised forecast generation would likely be unsuccessful, primarily due to the disjunction between the optimization goal (a function of predicted and true values) and the evaluation goal (a financial performance metric) \citep{Kovavcevic2022}. An alternative approach involves directly modeling a trading strategy using the labels from trade decisions. However, the efficacy of this system depends on the accuracy of the labeling algorithm, making it ultimately reliant on external factors. 
By merging forecast and allocation tasks, DRL provides a comprehensive framework for modeling trading strategies. Here, the agent directly optimizes the investor's utility, accounting for various side-effects such as commission and slippage -- aspects that are more challenging to integrate into conventional supervised learning algorithms \citep{Moody2001}. 

Financial literature builds trading strategies primarily relying on the historical observation of market data \citep{Moskowitz2012, Poterba1988}. Technical indicators, heuristic-based signals derived from market data, are employed in a forecast module, which is followed by a simplistic rule-based allocation module to create a trading strategy. Many of these strategies revolve around two key principles: momentum and mean-reversion. Momentum strategies \citep{Moskowitz2012} expect prices to keep their historical trends, while mean-reversion strategies \citep{Poterba1988} expect prices to revert to their long-term mean. However, a significant challenge lies in the fact that strategies constructed from arbitrary indicators often lack generalizability and tend to be effective only in specific market regimes. As such, these strategies may not perform well in intraday scenarios, primarily due to the heightened return variability at intraday granularity.

Recently, there has been an outburst in research on intraday price predictability across diverse markets \citep{Huddleston2023, Wen2022}. Results provide strong evidence that machine learning methods can effectively predict market returns within intraday horizons, beyond what can be explained by transaction costs \citep{Huddleston2023}. Building on this predictability, DRL methods create strategies based on state spaces constructed using diverse price-based features such as raw prices, returns, and technical indicators, with limited consideration for the agent's contextual information about its position relative to specific points in time.

In \citep{Deng2016}, the authors introduced a model called \emph{fuzzy deep direct reinforcement learning}. This model is comprised of three main components: a fuzzy learning preprocessing stage aimed at reducing uncertainty in noisy financial data, a recurrent neural network for feature representation, and an allocation module based on RL. The model makes decisions on a minute frequency. The input to the system are price changes of the last 45 minutes and the change to the previous 3 hours, 5 hours, 1 day, 3 days, and 10 days, forming a $50$--dimensional vector. Position, the unit amount of an asset owned, is appended to the input vector for the allocation module. Besides the position, no other contextual information is given to the model. The evaluation dataset involved three futures contracts (an equity index and two commodity assets) and covered a one-year period. The authors concluded that the RL framework is particularly well-suited for momentum strategies, showcasing consistent profits in markets with significant directional movements. 

In another study, a deep recurrent Q-network model that makes trading decisions in 15-minute intervals was proposed \citep{Huang2018}. This model assumes that positions can be held overnight. Universal state space from multiple asset returns, along with their lagged values, is constructed and forms a $198$-dimensional feature vector. State space also includes current time and position as features providing contextual information. The model was evaluated over a period of five years using foreign exchange securities.  

The authors in \citep{Si2017} proposed a DRL model consisting of fully connected (FC) layers for feature learning, followed by a combination of a long short-term memory (LSTM) layer and an FC layer used for decision-making. The model operates on a minute frequency, and its' state space features encompass the most recent $200$ asset returns. Even though the LSTM layer incorporates the sequentiality of decision-making, no contextual information is given. The model was evaluated on two months of data for three stock index futures in China. 

An extended deep Q-network (DQN) and asynchronous advantage actor-critic (A3C) models which operate on a minute frequency were proposed in \citep{Li2019}. These models use price data and technical indicators as price-based features. Additionally, they include the remaining cash and the Sharpe ratio from the previous time step as features providing contextual information. The authors evaluated the model on six assets, four stocks, and two equity futures, on roughly one year of data. Furthermore, they demonstrated that the actor-based A3C model exhibited superior performance compared to the value-based DQN. This distinction is attributed to the actor-based A3C operating directly in the policy space. 

A study conducted in \citep{Liu2020} proposes the \emph{imitative recurrent deterministic policy gradient} model which combines imitation learning \citep{Goluza2023} with a deterministic policy gradient, and operates on a 15-minute frequency. The model uses price data and technical indicators as price-based features and provides contextual information through the account profit feature. The authors evaluated the model using a dataset spanning up to one year, focusing on two equity futures.

The authors presented \emph{DeepScalper} in \citep{Sun2022}, a DRL model operating on a minute frequency using both macro-level and micro-level market data. Price data at different granularities is used to extract micro-level and macro-level market information. The model incorporates the current position, cash, and remaining time as features providing contextual information. It was evaluated across four fixed-income futures and two equity futures over three months. It's noteworthy that our proposed model does not necessitate order book data for constructing learning features, distinguishing it from their approach.

Out of existing studies, most models have predominantly utilized price-based features to construct intraday trading strategies, with momentum features being a prime example. These features come from a hypothesis that past performance is the main driver of asset returns, and they remain popular in financial literature because of their simplicity and consistent performance across different assets, geographical regions, and periods \citep{Moskowitz2012}. While price-based features provide valuable information, they do not offer insight into the context of the strategy's position, a crucial consideration given the sequential nature of trading tasks. The strategy could be improved by taking into account its position relative to specific points in time, such as assessing the remaining time in the trading day to mitigate overnight risks by selling assets before the market closes. We refer to features created in this way as positional features, and they encapsulate the strategy's position relative to both historical and future contexts. Moreover, existing studies evaluate models on a small sample of assets or relatively short periods, while having models with state spaces several orders of magnitude larger than our proposed model. This underscores the need for further exploration, both for robust strategy validation and an understanding of the strategy's behavior.

This paper introduces three contributions. Primarily, we present a novel DRL model for intraday trading strategies, which integrates positional features as a pivotal component. Moreover, we conduct a comprehensive evaluation of our proposed model over an extensive period, encompassing a diverse range of assets including commodities and foreign exchange securities. We also provide insights into the intraday trading activity of the model through the analysis of trades within specific time intervals.

The rest of the paper is organized as follows: In Section \ref{sec:problem_formulation}, we formally introduce the problem considered in this work. Section \ref{sec:methodology} provides an overview of the fundamental DRL concepts and detail each component of our DRL model. In Section \ref{sec:experimental_setup}, we discuss the dataset used, specifics of the DRL training process and environment, and explain the evaluation process. Section \ref{sec:results} presents the performance of the DRL trading strategy in comparison with benchmarks, describes trading activity, and highlights feature importance. Finally, in Section \ref{sec:conclusion}, we summarize the main findings and suggest potential directions for future research.

\section{Problem Formulation}\label{sec:problem_formulation}
Each asset has its limit order book (LOB), composed of limit orders that encapsulate all the information required to trade by representing the trader's desire to buy or sell \citep{Cont2011}. A limit order is defined by its components: side (bid/ask), size (quantity of units), limit price (the designated trading price), and submission time. The bid side represents buy orders, while the ask side represents sell orders. When a trader submits a market order, specifying the side and size, they are promptly matched with enough limit orders to meet their desired size at the best available price, resulting in the execution of a trade and the formation of a price. In the provided example LOB shown in Table~\ref{table:lob}, consider a sell market order with a size of $2000$. When triggered, the trader would execute the sale of $1000$ units for $100$, followed by the sale of another $1000$ units for $99.9$. This is because the sell market order matches with the existing bid orders in the limit order book, leading to these executions. 

\begin{table}[h!]
\caption{An example of a limit oder book}\label{table:lob}
\begin{tabular}{c c c} 
 \toprule
 \multicolumn{1}{c}{\textbf{Price}} & \multicolumn{1}{c}{\textbf{Size}} & \multicolumn{1}{c}{\textbf{Side}} \\ 
 \midrule
 99.9 & 3000 & Bid \\
 99.9 & 4000 & Bid \\
 100 & 1000 & Bid \\
 \hline
 100.1 & 5000 & Ask \\
 100.2 & 2000 & Ask \\
 100.3 & 1000 & Ask \\
 \bottomrule
\end{tabular}
\end{table}
To summarize the market dynamics, LOB is often sampled and aggregated into OHLCV (\emph{Open}, \emph{High}, \emph{Low}, \emph{Close}, \emph{Volume}) data, providing a summarized view of price movements within consistent time intervals, each defined by the duration $\Delta t$. For clarity, the ending of each interval is indexed by the discrete time step $t$. The \emph{Open} is represented by the first observed price $p_t^O$, while the \emph{High} and \emph{Low} correspond to the highest and lowest prices $p_t^H$ and $p_t^L$ during the sampling period, respectively. The \emph{Close} is the last observed price $p_t^C$. Additionally, \emph{Volume} denotes the total quantity of assets traded during that time interval. We use $\Delta t=1$ minute in our study, as this value represents intraday trading granularity.

We consider a trading strategy for a single asset that maintains flat positions by the end of the trading day to mitigate overnight risks and account for unpredictable events outside of market hours, which may significantly impact market movements. In DRL terms, we define a strategy as a policy $\pi$, a deterministic function that outputs the position $a_t = \pi(s_t)$ for each time step $t$ based on the state $s_t$ that encompasses historical information available up to that time step. The position $a_t \in \{-1,0,1\}$ refers to the unit amount of an asset owned. A negative position, known as a \emph{short} position, signifies that a strategy sells an asset that it does not own (known as \emph{selling short}) and speculates that the price of the asset will fall. In contrast, a positive position, known as a \emph{long} position, signifies that a strategy holds an asset with the expectation that its price will rise. Strategy's objective is the maximization of the logarithmic return, factoring in transaction costs denoted as $tx$:
\begin{equation}
\label{eq:return}
r_{t+1}^s = \log \left(\frac{p_{t}^C + a_t \cdot (p_{t+1}^{C} - p_{t}^{C}) - tx_{t+1}}{p_{t}^{C}}\right)
\end{equation}
This logarithmic return is included in the reward function of a DRL model. In line with this, we focus on designing a novel DRL model for intraday trading that integrates positional features in its state space, as well as empirically demonstrating their positive impact on model performance.

\section{Methodology}\label{sec:methodology}
In this section, we introduce the concept of the Markov decision process (MDP) and present the specifics of MDP for our DRL model. We also provide a detailed presentation of the algorithm at the core of our model -- the proximal policy optimization (PPO) algorithm.

\subsection{MDP Formulation}
Formally, an MDP is a tuple $\langle \mathcal{S}, \mathcal{A}, \mathcal{P}, \gamma, \mathcal{R}, p_0 \rangle$:
\begin{itemize}
        \item $\mathcal{S}$ is a set of states,
        \item $\mathcal{A}$ is a set of actions,
        \item $\mathcal{P} : \mathcal{S} \times \mathcal{A} \times \mathcal{S} \mapsto [0,1]$ is a transition probability matrix,
        \item $\gamma \in[0,1]$ is a discount factor,
        \item $\mathcal{R} : \mathcal{S} \times \mathcal{A} \times \mathcal{S} \mapsto \mathbb{R}$ is a reward function,
        \item $p_0$ is an initial state distribution.
\end{itemize}
At each time step $t$, the agent makes an action $a_t$ based on the environment's current state observation $s_t$. Performing an action $a_t$ in state $s_t$ determines the next state $s_{t+1}$ and the reward $r_{t+1}$. An episode (also called a trajectory) is a sequence of states and actions $\tau = (s_0, a_0, s_1, a_1, ...)$. Depending on the environment, the length of an episode is either fixed or it depends on the actions of the agent. The agent interacts with the environment until an episode is concluded, and the training process encompasses multiple episodes of interaction. MDP states have Markov property, meaning that the next state only depends on the current state and the current action, not on the history of all the states and actions that were taken before. Important concepts in DRL are value functions of state or state-action pairs. They represent the expected cumulative reward starting from the given state (or state-action pair) and acting according to a particular policy over all future time steps:

\begin{align}
    V^{\pi} (s) &= \mathbb{E}_\pi[R(\tau) | s_0 = s]\\
    Q^{\pi} (s, a) &= \mathbb{E}_\pi[R(\tau) | s_0 = s, a_0 = a]
\label{eq:v_opt}
\end{align}
We assume a finite-horizon setup, where the length of an episode is fixed. The expression $R(\tau) = \sum_{t}^{T} \gamma^t r_t$ represents the cumulative reward over the course of a trajectory, where $T$ is the time horizon, and $r_t$ represents the reward at time step $t$. $\gamma$ determines the present value of future rewards. In environments characterized by an infinite time horizon, $\gamma$ is typically less than $1$ to ensure mathematical convenience. Conversely, in finite-horizon environments, it is commonly set to 1.

The agent's behavior is determined by its policy $\pi$, which can either be a deterministic or a stochastic function mapping states to actions. The agent's primary goal is to find a policy $\pi^*$ that maximizes the expected cumulative reward over all future time steps:
\begin{equation}
    \pi^* = \argmax_{\pi_\theta} \mathbb{E}_{\tau \sim p_{\theta}(\tau)}[R(\tau)].
\label{eq:pi_opt}
\end{equation}
where $p_{\theta}(\tau)$ represents a parameterized probability of a trajectory which is induced by the parameterized policy $\pi_{\theta}$, and $\mathbb{E}_{\tau \sim p_{\theta}(\tau)}[R(\tau)]$ represents the expected value of cumulative rewards under that same policy $\pi_{\theta}$. Parameterized policy is a deep neural network in the DRL setup, and the final goal is to learn the network parameters of the optimal policy $\pi^*$. Any data-driven algorithm well suited to solving problems formulated in terms of an MDP and yielding an optimal policy can be considered an RL algorithm \citep{Sutton2018}. 
These algorithms take various forms: actor-based, directly optimizing the policy; value-based, optimizing value functions for each state or state-action pair, with the policy being derived indirectly from these estimates; or actor-critic, wherein both value and policy functions are concurrently learned to guide the agent's learning process \citep{Sutton2018}. 

\begin{figure}
  \centering
  \includegraphics[width=0.7\textwidth]{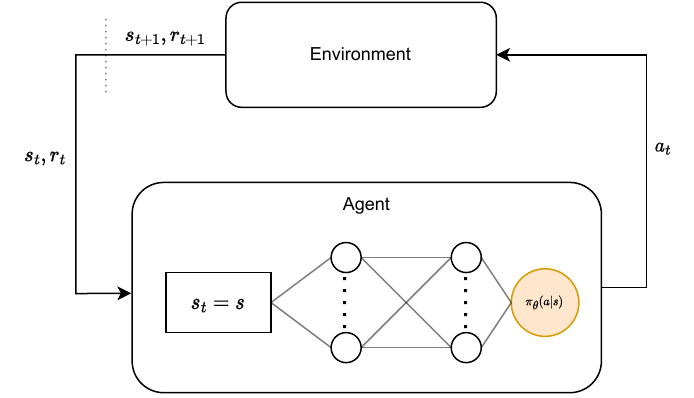}
  \caption{Sequential-decision making of the agent}
  \label{fig:drl_agent_env}
\end{figure}

A trading strategy aligns seamlessly with DRL's concept of a policy. In this scenario, states encompass historical information available up to a given time step, and actions correspond to the positions taken by the trading strategy. Finding the optimal policy $\pi^*$ translates to finding a strategy that optimizes the chosen reward, typically representing the investor's utility, whether it be focused on maximizing profitability or enhancing risk-adjusted metrics. Fig. \ref{fig:drl_agent_env} visualizes the sequential decision-making process of the agent for the trading strategy problem.

In this study, we establish an MDP framework to address the presented problem. Subsequently, we outline the constituents of the MDP, including the state, action, and reward components in the upcoming subsections.

\subsubsection{State Space}
As mentioned in Section \ref{sec:intro}, different features have been utilized to define state spaces. We incorporate price-based features but also introduce novel positional features that provide valuable information about the agent's positional context. LOB data is not used, rather only OHLCV data, as it's shown that the model does not depend on highly granular data to work. At each time step $t$, the agent observes a sparse state $\bm{s}_t \in \mathbb{R}^{13}$ comprising of two subsets of features:  positional and price-based.

Price-based features $\bm{z}_t \in \mathbb{R}^9$ incorporate the asset's past performance \citep{Moskowitz2012}. They include asset's past returns and technical indicators, features developed by economists to capture different properties of the financial time series \citep{Murphy1999}. These features are:
\begin{itemize}
    \item Returns over various lookback windows, specifically the last minute, five minutes, fifteen minutes, thirty minutes, and sixty minutes. These are calculated as:
    \begin{equation}
    r_{t-w,t}  = \frac{p_{t}^C - p_{t-w}^C}{p_{t-w}^C}, \, w \in \{1,5,15,30,60\}
    \end{equation}
    \item Relative strength index ($RSI$): measures the speed and magnitude of an asset's recent price changes to evaluate overvalued or undervalued conditions. Refer to \citep{Murphy1999} for detailed equations. A lookback window of $w=14$ is used.
    \item Average directional index ($ADX$): signals the presence of a strong trend in either direction. Refer to \citep{Gurrib2018} for detailed equations. A lookback window of $w=14$ is used.
    \item Ultimate oscillator ($ULTOSC$): measures the price momentum of an asset across multiple timeframes using weighted averages. Refer to \citep{Chan2016} for detailed equations. Lookback windows of $w_1=7, w_2=14, w_3=28$ are used.
    \item Williams \%R ($WILLR$): measures overbought and oversold conditions. Refer to \citep{Murphy1999} for detailed equations. A lookback window of $w=14$ is used.
\end{itemize}

Positional features $\bm{\rho}_t \in \mathbb{R}^4$ tend to capture the agent's positional context by providing information about position relative to specific points in time. These features are important for the sequentiality of the decision-making process. Within this feature group, the following features are encompassed:
\begin{itemize}
    \item Time left ($TL$): represents the remaining time in the current trading day, measured in minutes. It is calculated as:
    \begin{equation}
        TL = 360 - t  
    \end{equation}
    as there are $360$ time steps the agent takes during trading hours in one day. This presents the information about future context, assessing the remaining time to adapt the policy based on specific conditions in different periods of the day.
    \item Position ($POS$): the agent's current position. It can obtain three possible values $POS \in \{-1,0,1\}$.  
    \item Position return ($PR$): return between the time step of entering the current position, and the current time step, after transaction costs. If $t_p$ represents the time step of entering the current position, this feature is calculated as follows:
    \begin{align}
        PR_{t} &=  \frac{a_{t_p} \cdot (p_{t}^C - p_{t_p+1}^O) - tx_{t_p+1}}{p_{t_p+1}^O}\\
        tx_{t_p+1} &= \text{COM} \cdot p_{t_p+1}^O \cdot |a_{t_p}-a_{t_p-1}| \nonumber
    \end{align}
    The transaction costs incurred at time step $t_p$ are denoted as $tx_{t_p}$. These costs are explained in detail later in the section. 
    \item Daily return ($DR$): return between the initial time step of the trading day and the current time step, after transaction costs. Let $t=0$ represent the initial time step of the trading day. For each position change in the trading day up to the current time step, we define a historical position return $HPR$. This is similar to calculating historical $PR$ feature, as for the $i$-th position change, we denote the time step $t_{enter}^i$ of entering the $i$-th position, and the time step $t_{exit}^i$ of exiting the $i$-th position:
    \begin{align}
        HPR^i &=  \frac{a_{t_{enter}^i} \cdot (p_{t_{exit}^i+1}^O - p_{t_{enter}^i+1}^O) - tx_{t_{enter}^i+1}}{p_{t_{enter}^i+1}^O} \label{eq:hpr} \\
        tx_{t_{enter}^i+1} &= \text{COM} \cdot p_{t_{enter}^i+1}^O \cdot |a_{t_{enter}^i}-a_{t_{enter}^i-1}| \nonumber
    \end{align}
    Let $I$ be the total number of position changes up to the current time step. The daily return is calculated as: 
    \begin{equation}
        DR_{t} = \frac{\sum_{i=1}^{I-1} (p_{t_{enter}^i+1}^O \cdot HPR^i) + p_{t_p+1}^O \cdot PR_{t}}{p_{0}^O}
    \end{equation}
\end{itemize}
Notice that the last position change is the numerator of the current $PR$ feature. Positional features represent the agent's position relative to historical ($POS$, $PR$, $DR$) and future ($TL$) contexts. Finally, state space is formed by concatenating price-based and positional vectors $\bm{s}_t = [\bm{z}_t, \bm{\rho}_t]$.

\subsubsection{Action Space}

We design our agent to rely on market orders for trade execution. We also want him to be able to enter both sides of the market. To satisfy these presumptions, we formulate the action space as a discrete action space consisting of $a_t \in \{-1,0,1\}$. This action space also represents the position the agent needs to take in the asset, not the actual trading decisions of buying or selling the asset \citep{Zhang2020}. For example, if the position $a_{t-1}=0$ is followed by the position $a_t=1$, it implies that we need to buy $|a_{t} - a_{t-1}|=1$ unit of the asset. If the next position is $a_{t+1}=-1$, we would need to sell the asset we are holding, to exit the current position, and then enter the position $-1$ by selling short. Notice that in this situation the transaction costs will be doubled as we perform $|a_{t+1} - a_t| = 2$ transactions. This presents a solution to the risk management problem of position sizing, helping to prevent the accumulation of large positions.

\subsubsection{Reward Function}

To satisfy the additive rewards required by the DRL optimization goal, we define the reward function as the logarithmic return at each time step:

\begin{align}
r_{t+1} &= \log \left(\frac{p^{\text{exec}} + a_{t} \cdot (p_{t+1}^{C} - p^{\text{exec}}) - tx_{t+1}}{p^{\text{exec}}}\right)\\
&tx_{t+1} = \text{COM} \cdot p_{t+1}^{O} \cdot |a_{t} - a_{t-1}|\\
&p^{\text{exec}} = \begin{cases}
    p_{t+1}^{O},& \text{if } |a_{t} - a_{t-1}|\neq 0\\
    p_{t}^{C},& \text{otherwise}
\end{cases}
\end{align}
The transaction costs $tx_t$ include commission, slippage, and market impact. 
Commission cost estimates ($\text{COM}$) are expressed in basis points (BP), where $1$BP equals $0.01\%$. These estimates determine the percentage of the traded price used as a commission. Slippage is conservatively approximated by assuming that trades execute at the \emph{Open} price $p_{t+1}^{O}$ of the next time step, which is not known at the time of decision-making. Therefore, $p^{\text{exec}}$ denotes that price if the agent changes positions, and the previous time step's \emph{Close} price $p_{t}^{C}$ if no trade is executed. Given that our strategy involves trading small positions of one unit, we assume that its market impact is negligible, meaning it does not significantly influence market movements.

\subsection{Proximal Policy Optimization}

In this study, we use the proximal policy optimization (PPO) algorithm \citep{Schulman2017},  a model-free actor-critic DRL algorithm. The model-free nature of PPO implies that it does not necessitate learning the dynamics or transition probabilities of the trading environment \citep{Liu2021}.  Instead, PPO operates directly within the policy space, harnessing the gradient of policy performance -- specifically, the anticipated cumulative reward (Eq. \ref{eq:pi_opt}) -- to refine and optimize the trading policy \citep{Yang2020}. The algorithm relies on the advantage function $A(s, a)$, defined as the difference between the state-action value $Q(s, a)$ and the state value $V(s)$ for a given state $s$ and action $a$. This function quantifies the additional reward that the agent could obtain in state $s$ by taking action $a$ compared to the average reward of actions at that state. In essence, it guides the agent in increasing the probabilities of actions that lead to higher rewards and reduces the variance of the policy gradient updates. 

Despite being an on-policy algorithm, wherein the estimated policy is utilized for both sampling and learning, it possesses a distinct advantage. The incorporation of a clipped objective function prevents excessively large policy updates between consecutive steps. This not only ensures stability but also facilitates the efficient reuse of experience samples, allowing for multiple gradient steps on the same mini-batch of experience. Furthermore, the use of multiple actors concurrently collecting experience adds another layer of efficiency to the process. PPO's conservative policy improvement, along with its use of first-order optimization, strikes a balance between simplicity, sample efficiency, and robustness. These qualities make it a crucial component of our proposed agent.

The PPO algorithm is presented in detail in Algorithm \ref{alg:PPO}. In our implementation, the final objective (Eq. \ref{eq:total_loss}) consists of a clipping objective (Eq. \ref{eq:clip_loss}) and a value function loss (Eq. \ref{eq:vf_loss}). The clipping objective prevents excessively large policy updates by introducing a surrogate objective that clips the probability ratio $\rho_t$ between the old and new policies. $\hat{\mathbb{E}}_t$ denotes the empirical expectation over a finite batch of samples. The value function loss is a mean-squared loss used for learning the state value function $V_{\theta}(s)$ \citep{Li2023}. It is implemented as a neural network that shares all layers with the policy $\pi_\theta$, except the final one, which outputs the state value instead of the action probabilities. The value function is utilized in the computation of the advantage function. We estimate the advantage function using the exponentially-weighted generalized advantage estimation (GAE) method \citep{Schulman2015}.

\begin{algorithm}
\caption{Proximal Policy Optimization}\label{alg:PPO}
\begin{algorithmic}[1]
\State \textbf{for} iteration $i=1,2,..$ \textbf{do}:
\State \quad \textbf{for} actor $a=1$ to $N$ \textbf{do}:
\State \quad \quad Execute policy $\pi_{\theta_{i-1}}$ for $B$ time steps
\State \quad \quad Calculate advantage function estimates: $\hat{A}_1, \hat{A}_2, ..., \hat{A}_B$
\State \quad \textbf{endfor}
\State \quad Update $\theta$ using the gradient ascent with $\nabla L_{\theta}$, with $K$ epochs and minibatch \quad size $M \leq NB$

\begin{align}
    L_t(\theta) &= \hat{\mathbb{E}}_t[L_t^{\text{CLIP}}(\theta) - c \, L_t^{\text{VF}}(\theta)] \label{eq:total_loss}\\
    L_t^{\text{CLIP}}(\theta) &= \hat{\mathbb{E}}_t[\text{min}(\rho_t(\theta) \hat{A}_t, \text{clip}(\rho_t(\theta), 1 - \epsilon, 1 + \epsilon)\hat{A}_t)] \label{eq:clip_loss}, \\
    &\text{where } \rho_t(\theta) = \frac{\pi_{\theta_i(a_t | s_t)}}{\pi_{\theta_{i-1}(a_t | s_t)}} \\
    L_t^{\text{VF}}(\theta) &= (V_{\theta}(s_t) - \hat{V}(s_t))^2 \label{eq:vf_loss}
\end{align}
\State \textbf{endfor}
\State \textbf{return} optimized policy parameters $\theta$
\end{algorithmic}
\end{algorithm}

\section{Experimental Setup}\label{sec:experimental_setup}

In this section, we provide an overview of the data used for training and evaluation, details of the environment and agent training procedures, and introduce the evaluation metrics and benchmark strategies.

\subsection{Data and Preprocessing}

We use a continuous futures dataset composed of OHLCV data sampled at $\Delta t=1$ min. This continuous futures price series is a time series of futures prices created by combining the active front month contracts. To ensure consistency for testing, we adjust prices to account for price jumps on contract roll dates. This results in a suitable time series for evaluation \citep{Vojtko2020}. 

One of the primary challenges in financial literature is avoiding data snooping, which refers to seemingly statistically significant results from unrealistic setups and overfitting \citep{Bailey2014}. Given the various potential pitfalls (such as the survivorship bias in asset selection), we address this by isolating the first year of data -- never used for testing purposes -- specifically for conducting data analysis to inform our choice of assets for trading. 
We initially select assets with liquid trading hours between 09:30 and 17:00 (ET). This decision is informed by historical data from Interactive Brokers and it ensures that our trading aligns with multiple assets without introducing future bias. Subsequently, we assess the daily volume of traded prices and address any missing data. In cases of missing samples, we replace OHLC price information with the last traded close price, and we assign a value of zero for traded volume data. Additionally, we filter out assets with multiple days where less than 1000 units of contracts were traded. This is crucial, as in illiquid markets, insufficient order flow could lead to suboptimal trade execution prices, a situation we aim to avoid.

In the end, we choose four foreign exchange futures and six commodity futures which are listed in Table \ref{tab:assets}. The dataset spans from 2012 to the end of 2021. Each asset is driven by its unique set of underlying factors. For commodities, prices are influenced by supply and demand dynamics formed by hedgers and speculators, geopolitical events, and industrial usage. In contrast, even though there are some similar factors such as geopolitical events, foreign exchange futures are influenced by factors such as central bank policies, trade balances, and political stability, resulting in less volatile price changes and more persistent price trends. 

For training and testing the model, we employ a rolling window time series approach \citep{Kolm2023}. Each \emph{roll} encompasses a training period of one year followed by a testing period of four months. The last month of the training period serves as a separate validation set for early stopping. During each roll, the model is exclusively trained on data from the corresponding training set, without incorporating any previous data. This testing methodology yields a total of nine years of testing data, covering the period from 2013 to the end of 2021, distributed across 27 \emph{rolls} in total.

\begin{table}[h]
\caption{Futures contracts used in experiments denominated in US Dollars}
\begin{tabular}{llc}
\toprule
\multicolumn{1}{c}{\textbf{Identifier}} & \multicolumn{1}{c}{\textbf{Asset}} & \textbf{Asset class} \\
\midrule
GC                                      & Gold                               & Commodity \\
HG                                      & Copper                             & Commodity \\
SI                                      & Silver                             & Commodity \\
PL                                      & Platinum                           & Commodity \\
CL                                      & Crude Oil WTI                      & Commodity \\
NG                                      & Henry Hub Natural Gas              & Commodity \\
A6                                      & Australian dollar                  & Foreign Exchange \\
B6                                      & British Pound                      & Foreign Exchange \\
E1                                      & Swiss Franc                        & Foreign Exchange \\
MP                                      & Mexican Peso                       & Foreign Exchange    
\end{tabular}
\label{tab:assets}
\end{table}

\subsection{Environment Details}
Our environment simulates a single trading day, corresponding to a single episode. We choose to include only the liquid trading hours, which fall between 09:30 and 17:00 (ET). As there is a significant amount of volatility around the beginning and the end of these liquid trading hours, explained by the opening and closing of closely linked markets rather than the underlying price formation process, the agent does not trade during the initial and concluding half-hour periods. Additionally, to construct features for our state space, the agent requires a one-hour lookback window of intraday data, hence the agent does not trade in the first hour of the liquid trading hours. Under these conditions, the agent initiates trading at 10:31 and concludes by liquidating all open positions at 16:31, interacting with the environment across a fixed horizon of $T=360$ time steps.

\subsection{Training Details}
Before training our model, it is crucial to normalize the state space features, ensuring they are appropriately scaled for training. The specific feature value domains and normalization techniques can be found in Table \ref{tab:features}. The action space does not require normalizing as it is not continuous. 

The actor and critic neural networks are designed as multi-layered perceptrons (MLP), with two hidden layers consisting of 128 and 64 units respectively, and ReLU as activation functions. The learning rate is set to $\alpha = 0.0001$. The parameter $\lambda = 0.95$ accounts for balancing bias and variance for the GAE, while $c = 0.5$ is the coefficient for the value function loss in the total loss computation. We use the Adam optimizer \citep{Kingma2014} for optimizing the network parameters. A minibatch size of $M=64$ represents the sample size in gradient ascent, and a batch size of $B=832$ determines the size of the experience replay buffer.  The number of actors working in parallel is $N=3$. For this fixed-horizon intraday trading problem, the discount factor $\gamma$ is set to $1$. A commission term of $\text{COM}=0.08\text{BP}$ is used in the reward function. This term serves as a regularizer against excessive position changes. After completing one epoch of training on each day within the training period, the agent's performance on the validation set is assessed. If the total reward on the validation set does not show improvement over five consecutive epochs, the training process is halted. This application of early stopping, as well as having a neural network with a small number of parameters, serves as a mechanism to address potential overfitting to noise in the training data. The list of all hyperparameters is given in Table \ref{tab:hp}.

\begin{table}
\caption{State space features and normalization techniques}\label{tab:features}
\begin{tabular}{lcr}
\toprule
\multicolumn{1}{c}{\textbf{Feature}} & \multicolumn{1}{c}{\textbf{Values}} & \multicolumn{1}{c}{\textbf{Normalization}} \\
\midrule
$TL$   & $\in [0,359]$ & Min-max scaling to $[-1,1]$ \\
$POS$ & $\in \{-1,0,1\}$ & \multicolumn{1}{c}{-} \\
$PR$ & $\in \mathbb{R}$ & \begin{tabular}[c]{@{}r@{}}Standardardized with running $\hat{\mu}, \hat{\sigma}$ \\ estimated from the lookback window of $100$ episodes\end{tabular} \\
$DR$ & $\in \mathbb{R}$ & \begin{tabular}[c]{@{}r@{}}Standardardized with running $\hat{\mu}, \hat{\sigma}$ \\ estimated from the lookback window of $100$ episodes\end{tabular} \\
\hline
$\{1,5,15,30,60\}$--minute returns & $\in \mathbb{R}$ & \begin{tabular}[c]{@{}r@{}}Standardardized with running $\hat{\mu}, \hat{\sigma}$ \\ estimated from the lookback window of $5$ days\end{tabular}   \\
$RSI$ & $\in [0,100]$ & Min-max scaling to $[-1,1]$ \\
$ADX$ & $\in [0,100]$ & Min-max scaling to $[-1,1]$ \\
$ULTOSC$ & $\in [0,100]$ & Min-max scaling to $[-1,1]$ \\
$WILLR$ & $\in [0,100]$ & Min-max scaling to $[-1,1]$ 
\end{tabular}
\end{table}

\begin{table}
\caption{Hyperparameter values}\label{tab:hp}
\begin{tabular}{lllllllllll}
\toprule
\multicolumn{1}{c}{\textbf{$\alpha$}} & \multicolumn{1}{c}{\textbf{Optimizer}} & \multicolumn{1}{c}{\textbf{$M$}} & \multicolumn{1}{c}{\textbf{$B$}} & \multicolumn{1}{c}{\textbf{$N$}} & \multicolumn{1}{c}{\textbf{Layer units}} & \multicolumn{1}{c}{\textbf{Activation}} & \multicolumn{1}{c}{\textbf{$\lambda$}} & \multicolumn{1}{c}{\textbf{$c$}} & \multicolumn{1}{c}{\textbf{$\gamma$}} & \multicolumn{1}{c}{\textbf{COM}} \\
\midrule
0.0001 & Adam & 64 & 832 & 3 & (128,64) & ReLU & 0.95 & 0.5 & 1 & 0.08BP
\end{tabular}
\end{table}

\subsection{Evaluation}
Trading strategies are evaluated based on multiple performance metrics. These metrics include profitability and risk-adjusted metrics, with the most important one being the Sharpe ratio. To ensure comparability across strategies of varying scales, these metrics are often annualized.  As we deal with intraday trading, we calculate daily returns for each trading day in the testing period and annualize the metrics based on these daily returns. To investigate the various properties of our strategy, we include the following metrics \citep{Zhang2020}:
\begin{itemize}
    \item $\hat{\mathbb{E}}(r)$: annualized mean return ($252 \cdot \hat{\mathbb{E}}_{\text{daily}}(r)$)
    \item STD: annualized standard deviation of daily returns ($\sqrt{252} \cdot \text{STD}_{\text{daily}}$)
    \item DD: annualized downside deviation, represents the annualized standard deviation of negative daily returns
    \item MDD: maximum drawdown, represents how much strategy loses at its worst from its highest point before it starts recovering (worst-case scenario)
    \item Sharpe: annualized Sharpe ratio, represents the return earned in relation to the amount of risk taken ($\hat{\mathbb{E}}(r)/\text{STD}$)
    \item Sortino: annualized Sortino ratio, similar to Sharpe ratio except for including downside devation instead of standard deviation ($\hat{\mathbb{E}}(r)/\text{DD}$)
    \item Calmar: annualized Calmar ratio, similar to Sortino ratio except for including maximum drawdown instead of downside deviation ($\hat{\mathbb{E}}(r)/\text{MDD}$)
    \item \%(+)Rets.: percentage of positive daily returns
    \item $\frac{\hat{\mathbb{E}}_{\text{daily}^+}(r)}{\hat{\mathbb{E}}_{\text{daily}^-}(r)}$: the ratio between the mean positive and negative daily returns
\end{itemize}

To further assess the performance of our agent, we compare him against well-established passive trading strategy benchmarks. This allows us to determine whether our active intraday DRL agent strategy brings any performance improvements. Among the passive baseline strategies we incorporate are \emph{Buy \& Hold} and \emph{Sell \& Hold}. These strategies involve initiating either long ($1$) or short ($-1$) positions at the start of the testing period and maintaining them until the conclusion. In addition, considering that our state space is constructed from price-based features, we introduce another baseline strategy centered around momentum. This strategy leverages returns from the previous month. Specifically, at the start of each trading day, the agent assumes a long position if the return is positive, and a short position if it is not.

\section{Results and Discussion}\label{sec:results}
In this section, our DRL agent is evaluated based on the previously described methodology. Additionally, we provide insights into the agent's intraday trading activity during the evaluation process. The return of the agent $r^a$ between two time steps in test period for a specific asset is computed as follows:

\begin{equation}
\label{eq:ret_asset}
r_{t,t+1}^a  = \frac{a_{t} \cdot(p_{t+1}^{O} - p_{t}^{O}) - \text{COM} \cdot p_{t}^{O} \cdot |a_{t} - a_{t-1}|}{p_{t}^{O}}
\end{equation}
The commission used in the evaluation corresponds to the commission used in training. We use minute returns to construct cumulative return time series (Eq. \ref{eq:cum_ret}), often referred to as the PnL (Profit and Loss) curve. This curve provides a comprehensive view of the model's performance over time and different market conditions.
\begin{equation}
\label{eq:cum_ret}
r_{t}^{\text{cum}}  = \prod_{i=0}^{t-1} (1 + r_{i,i+1}^a) - 1
\end{equation}
We calculate daily returns and report the mentioned performance metrics. Table \ref{tab:sharpe_assets}. shows the annualized Sharpe ratio of the strategies for each asset throughout the test period. Our DRL agent demonstrates notable performance, outperforming the benchmarks in most assets and yielding high Sharpe ratios after accounting for realistic transaction costs. Specifically, in the commodities market, the agent outperforms the benchmarks for each asset, with Platinum being the top performer. This is particularly encouraging given the diverse nature of these commodities, which encompass sectors such as metals and energy. On the other hand, foreign exchange assets yield comparatively lower results, though they still produce positive Sharpe ratios. Our model outperforms the benchmarks in two out of four foreign exchange assets. For the other two assets, the Sell \& Hold strategy proves to be the most effective, with our model achieving competitive results. This can be attributed to the persistent negative trend observed throughout the test period. Despite the inherent volatility in prices, our model, in certain instances, may not capture enough inefficiencies to overcome the prevailing negative trend.
\begin{table}[h]
\caption{Sharpe ratio of strategies for each asset}\label{tab:sharpe_assets}
\begin{tabular}{lrrrr}
\toprule
{} &  \multicolumn{1}{c}{\textbf{DRL}} &  \multicolumn{1}{c}{\textbf{Buy \& Hold}} &  \multicolumn{1}{c}{\textbf{Sell \& Hold}} &  \multicolumn{1}{c}{\textbf{Momentum}} \\
\midrule
GC & \textbf{0.272} &       0.081 &       -0.076 &         -0.750 \\
HG & \textbf{1.479} &       0.138 &       -0.150 &          0.061 \\
SI & \textbf{1.293} &      -0.044 &        0.045 &         -0.331 \\
PL & \textbf{3.812} &      -0.156 &        0.135 &         -0.276 \\
CL & \textbf{0.791} &       0.206 &       -0.365 &          0.073 \\
NG & \textbf{1.068} &      -0.108 &        0.113 &          0.248 \\
\hline
A6 & 0.249 &      -0.285 &        \textbf{0.278} &          0.239 \\
B6 & 0.048 &      -0.232 &        \textbf{0.230} &         -0.198 \\
E1 & \textbf{0.545} &      -0.119 &        0.119 &         -0.595 \\
MP & \textbf{0.327} &      -0.048 &        0.030 &         -0.118 \\
\bottomrule
\end{tabular}
\end{table}

\begin{figure}[h]
  \centering
  \includegraphics[height=0.5\textheight,width=1\textwidth]{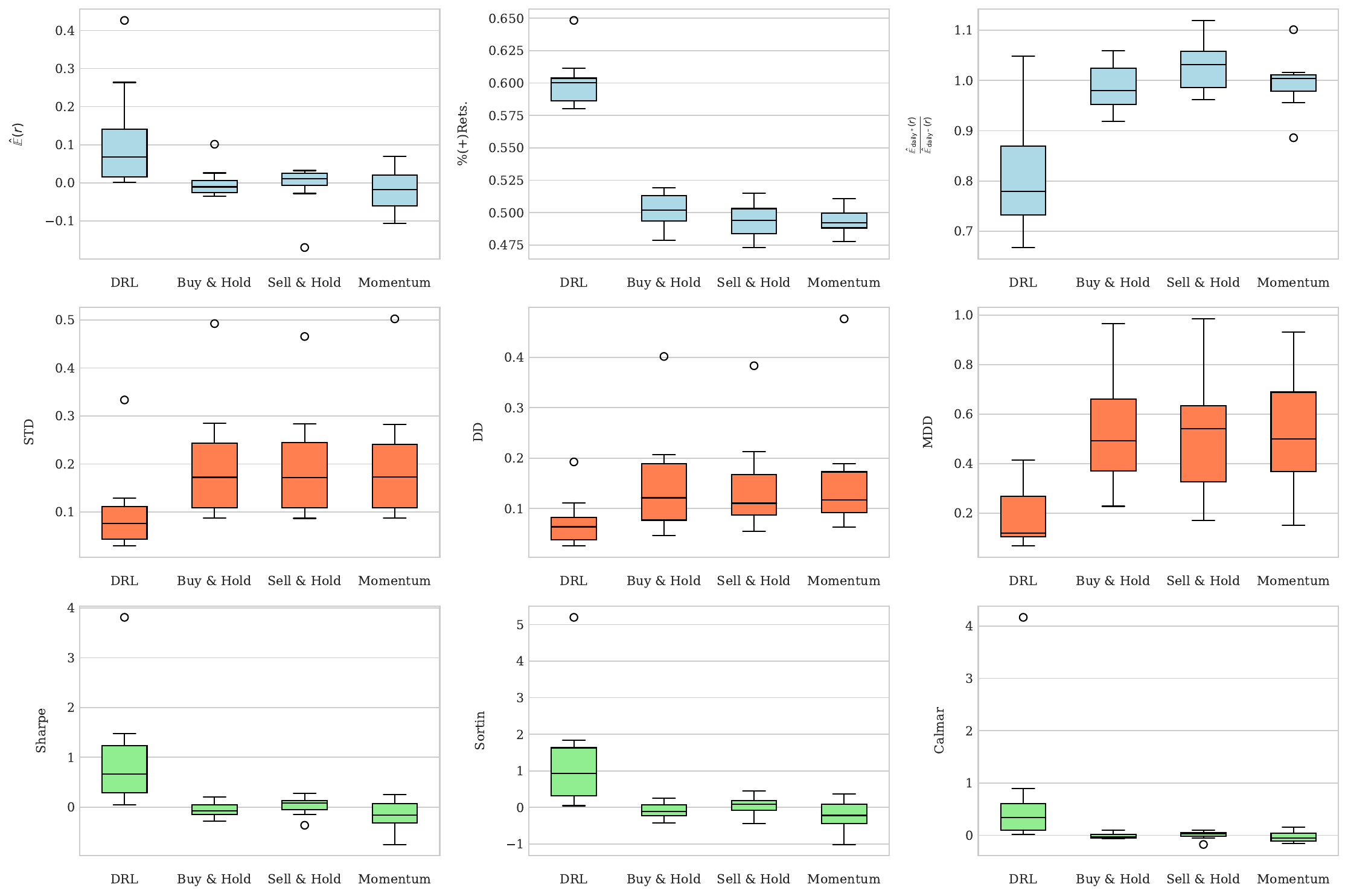}
  \caption{Performance metrics for different strategies}
  \label{fig:boxplot}
\end{figure}

Fig. \ref{fig:boxplot} uses box plots to visually present the performance metrics for each asset across different strategies. The DRL agent demonstrates notable performance in terms of annualized expected return. Additionally, the agent exhibits a higher proportion of positive daily returns. However, it's worth noting that the negative daily returns tend to be more substantial, attributed to the $\frac{\hat{\mathbb{E}}_{\text{daily}^+}(r)}{\hat{\mathbb{E}}_{\text{daily}^-}(r)}$ metric being below one for the majority of assets. 
Upon closer examination of risk metrics (STD, DD, MDD), it becomes evident that the DRL agent generates less volatile daily returns. It's important to emphasize that this reduction in volatility is not solely a result of the modeling, but rather a consequence of the fact that a substantial portion of the risk arises from overnight returns, to which all strategies, except DRL, are exposed.
Lastly, the evaluation of risk-adjusted metrics (Sharpe, Sortino, Calmar) reveals that our agent consistently outperforms benchmarks. These results also imply that our model is able to deliver a positive return relative to the level of risk assumed. Overall, these findings affirm the consistent performance of our agent across a diverse range of assets.

\begin{figure}[p]
  \centering
  \includegraphics[height=0.99\textheight, width=1\textwidth]{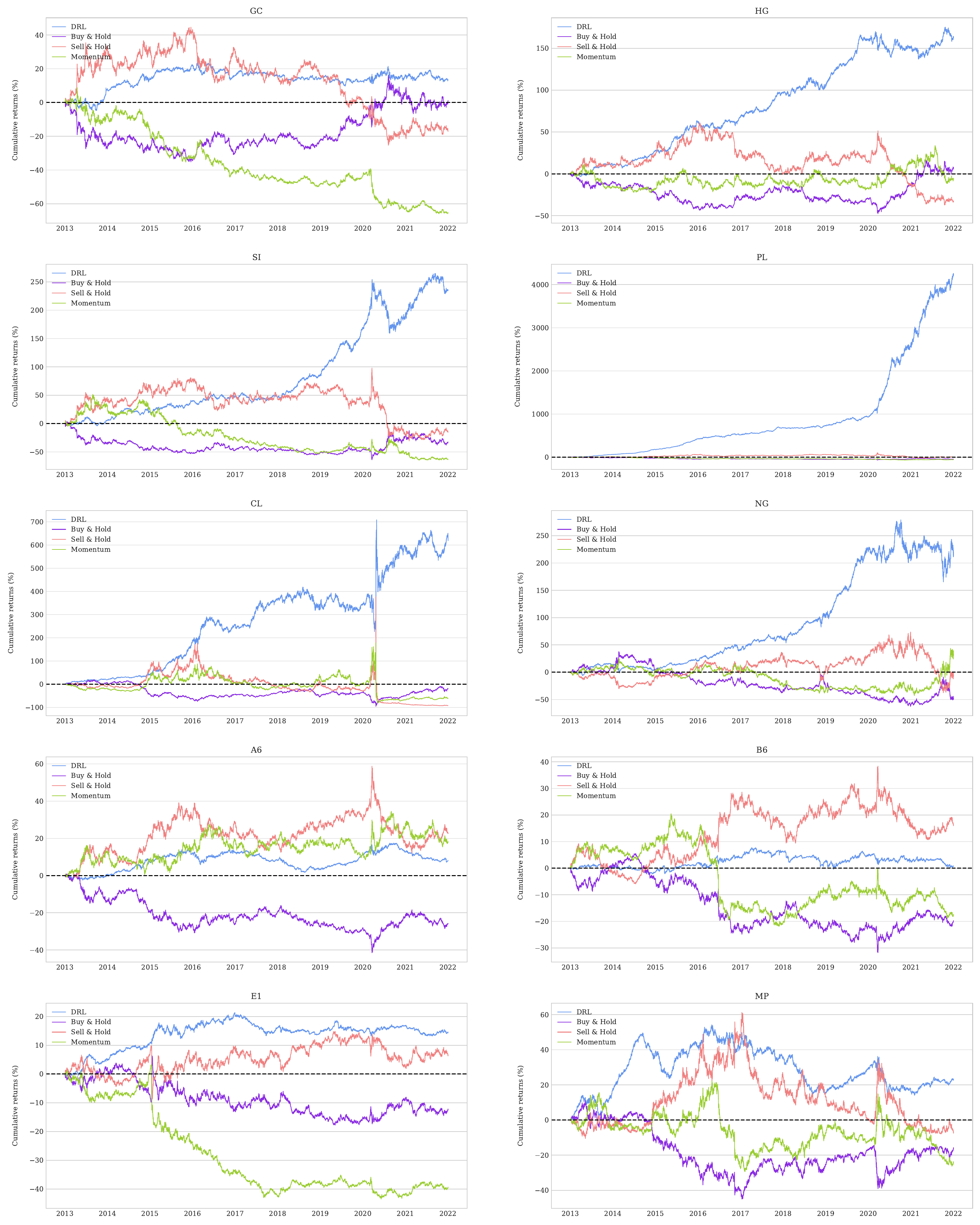}
  \caption{Cumulative returns for each asset}
  \label{fig:asset_pnl}
\end{figure}

\begin{figure}[h]
  \centering
  \includegraphics[height=0.5\textheight, width=1\textwidth]{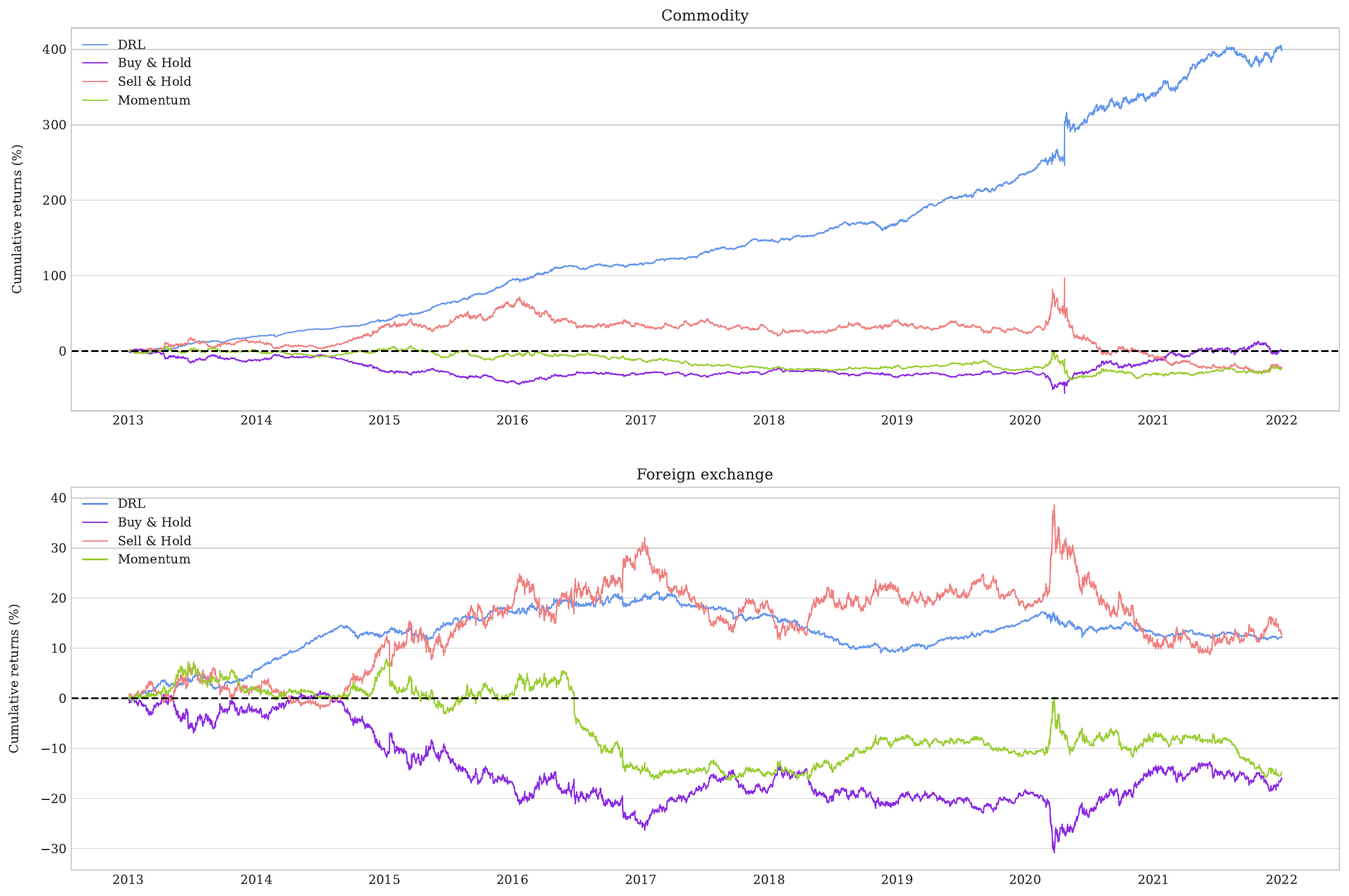}
  \caption{Cumulative returns for portfolios constructed for each asset class}
  \label{fig:portfolio_pnl}
\end{figure}

In Fig. \ref{fig:asset_pnl}, we present the PnL curve for each asset. This visual representation offers a clear overview of the performance of our agent across all assets. Additionally, we construct equal-weighted portfolios with distinct asset classes from our dataset: commodities and foreign exchange assets. The return of each portfolio, comprised of $N$ assets and their returns given by Eq. \ref{eq:ret_asset}, at every minute interval $\{t,t+1\}$ is expressed as follows:
\begin{equation}
r_{t,t+1}^{\,p} = \frac{1}{N} \sum_{i=1}^{N} r_{t,t+1}^{a,\,(i)} 
\end{equation}
Fig. \ref{fig:portfolio_pnl} illustrates the PnL curves of portfolios for each strategy. The performance metrics of these portfolios are presented in Tables \ref{tab:pnl_comm} and \ref{tab:pnl_fx}. Building upon our earlier analysis of individual asset performance, it is anticipated that portfolios of assets should generate even better results due to diversified risk. We observe the stable returns generated by our agent in the commodity asset class. Despite the agent's comparatively lower performance in the foreign exchange asset class, he still exhibits a relatively high Sharpe ratio compared to benchmarks and consistently provides stable returns even in a down-trending market throughout the testing period. This affirms the versatility and effectiveness of our agent across different asset classes and market regimes.
\begin{table}[h]
\caption{Performance metrics of commodity portfolios}\label{tab:pnl_comm}
\begin{tabular}{lrrrrrrrrr}
\toprule
{} & \multicolumn{1}{c}{\textbf{$\hat{\mathbb{E}}(r)$}} & \multicolumn{1}{c}{\textbf{STD}} & \multicolumn{1}{c}{\textbf{DD}} & \multicolumn{1}{c}{\textbf{MDD}} & \multicolumn{1}{c}{\textbf{Sharpe}} &  \multicolumn{1}{c}{\textbf{Sortino}} &  \multicolumn{1}{c}{\textbf{Calmar}}  \\
\midrule
DRL          &  \textbf{0.181} & \textbf{0.066} & \textbf{0.040} & \textbf{0.064} &   \textbf{2.759} &    \textbf{4.494} &   \textbf{2.847}\\
Buy \& Hold    &  0.014 & 0.156 & 0.116 & 0.575 &   0.090 &    0.121 &   0.025\\
Sell \& Hold   & -0.016 & 0.156 & 0.105 & 0.645 &  -0.105 &   -0.156 &  -0.025\\
Momentum & -0.020 & 0.140 & 0.109 & 0.421 &  -0.143 &   -0.183 &  -0.047 \\
\bottomrule
\end{tabular}
\end{table}
\begin{table}[h]
\caption{Performance metrics of foreign exchange portfolios}\label{tab:pnl_fx}
\begin{tabular}{lrrrrrrrrr}
\toprule
{} & \multicolumn{1}{c}{\textbf{$\hat{\mathbb{E}}(r)$}} & \multicolumn{1}{c}{\textbf{STD}} & \multicolumn{1}{c}{\textbf{DD}} & \multicolumn{1}{c}{\textbf{MDD}} & \multicolumn{1}{c}{\textbf{Sharpe}} &  \multicolumn{1}{c}{\textbf{Sortino}} &  \multicolumn{1}{c}{\textbf{Calmar}}  \\
\midrule
DRL          &  0.013 & \textbf{0.025} & \textbf{0.021} & \textbf{0.102} &   \textbf{0.520} &    \textbf{0.629} &   \textbf{0.130}\\
Buy \& Hold    & -0.016 & 0.078 & 0.053 & 0.319 &  -0.208 &   -0.302 &  -0.051\\
Sell \& Hold   &  \textbf{0.016} & 0.078 & 0.050 & 0.216 &   0.208 &    0.322 &   0.075\\
Mom (monthly) & -0.016 & 0.065 & 0.052 & 0.226 &  -0.242 &   -0.306 &  -0.070\\
\bottomrule
\end{tabular}
\end{table}

We now delve deeper into intraday patterns and extract relevant trading statistics. A single trade is characterized by the initiation of either long or short positions, followed by their subsequent exit. Table \ref{tab:trad_stats} shows the trading statistics of the agent for each asset. The win rate represents the percentage of positive return trades the agent has made. Terms $\hat{\mathbb{E}}_{\text{tr}^+}(r)$ and $\hat{\mathbb{E}}_{\text{tr}^-}(r)$ represent the mean positive and negative trade returns, respectively. The term $\hat{\mathbb{E}}_{\text{tr}}(r)$ represents the expected return on a trade, calculated using the win rate, mean positive trade returns, and mean negative trade returns. Lastly, we evaluate the mean duration of a trade, expressed in minutes. The analysis reveals that, in most cases, the agent secures a higher mean positive return per trade compared to the mean negative return, with a win rate hovering around 50\%. This pattern suggests that the agent's accurate assessments lead to more substantial gains, while its misjudgments result in relatively modest losses.
\begin{table}[h!]
\caption{Trading statistics of the agent}\label{tab:trad_stats}
\begin{tabular}{lrrrrrc}
\toprule
 & \multicolumn{1}{c}{\textbf{Win rate}} & \multicolumn{1}{c}{\textbf{$\hat{\mathbb{E}}_{\text{tr}^+}(r)$}} & \multicolumn{1}{c}{\textbf{$\hat{\mathbb{E}}_{\text{tr}^-}(r)$}} & \multicolumn{1}{c}{\textbf{$\frac{\hat{\mathbb{E}}_{\text{tr}^+}(r)}{\hat{\mathbb{E}}_{\text{tr}^-}(r)}$}} & \multicolumn{1}{c}{\textbf{$\hat{\mathbb{E}}_{\text{tr}}(r)$}} & \multicolumn{1}{c}{\textbf{Avg. Duration}} \\
 \midrule
GC & 55.22 \% & 0.027 \% & 0.032 \% & 0.834 & 0.0004 \% & 4.47 min\\
HG & 45.73 \% & 0.033 \% & 0.027 \% & 1.245 & 0.0007 \% & 3.19 min\\
SI & 43.63 \% & 0.048 \% & 0.036 \% & 1.342 & 0.0008 \% & 3.22 min\\
PL & 51.88 \% & 0.046 \% & 0.044 \% & 1.034 & 0.0025 \% & 3.65 min\\
CL & 50.86 \% & 0.077 \% & 0.076 \% & 1.005 & 0.0015 \% & 2.69 min\\
NG & 47.29 \% & 0.051 \% & 0.044 \% & 1.154 & 0.0008 \% & 2.79 min\\
A6 & 46.34 \% & 0.023 \% & 0.019 \% & 1.196 & 0.0003 \% & 4.95 min\\
B6 & 52.59 \% & 0.019 \% & 0.020 \% & 0.929 & 0.0003 \% & 6.58 min\\
E1 & 49.85 \% & 0.021 \% & 0.020 \% & 1.047 & 0.0004 \% & 6.83 min\\
MP & 40.52 \% & 0.046 \% & 0.031 \% & 1.493 & 0.0003 \% & 3.64 min\\
\bottomrule
\end{tabular}
\end{table}

\begin{figure}
  \centering
  \includegraphics[height=0.3\textheight, width=1\textwidth]{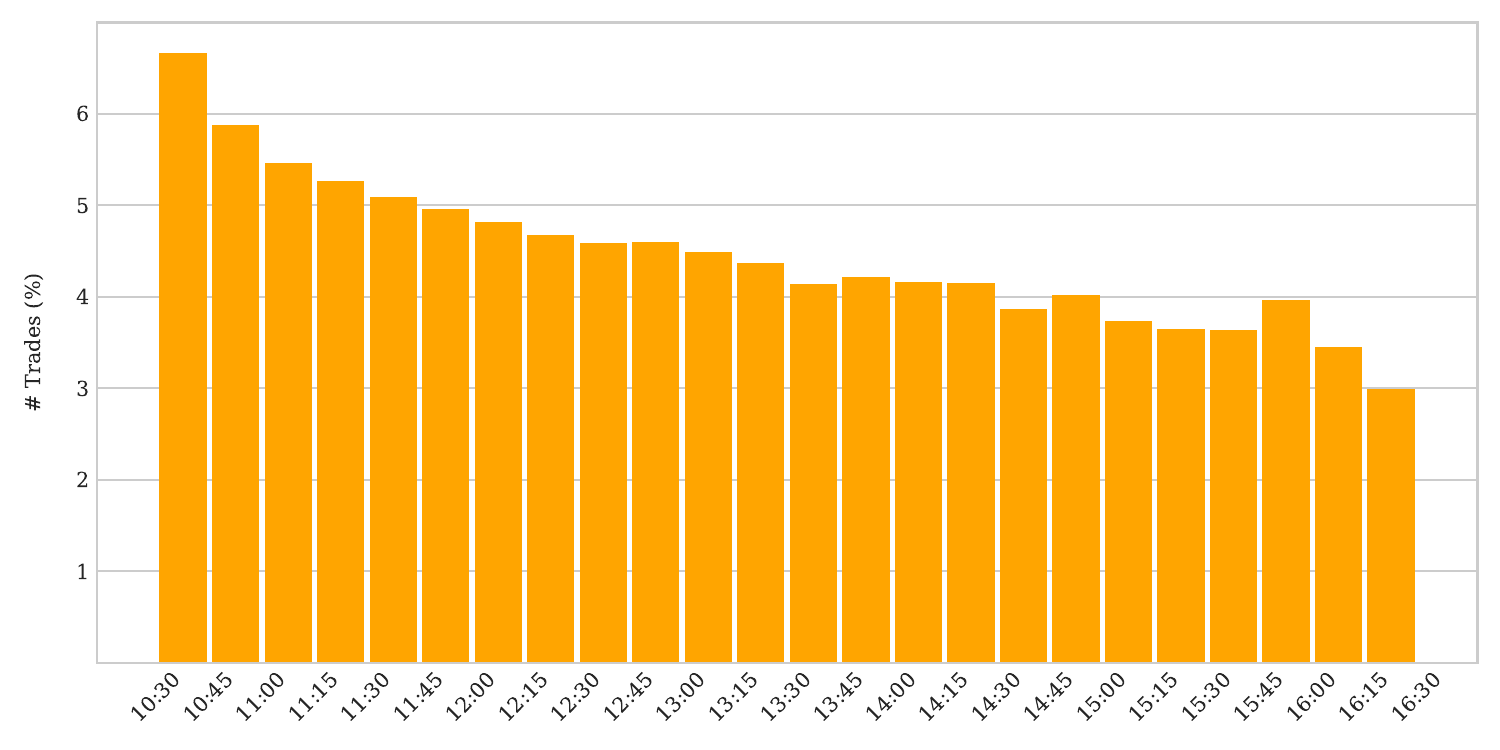}
  \caption{Percentage of trades initialized during specific time intervals}
  \label{fig:intraday_trades}
\end{figure}

\begin{figure}[t]
  \centering
  \includegraphics[height=0.3\textheight, width=1\textwidth]{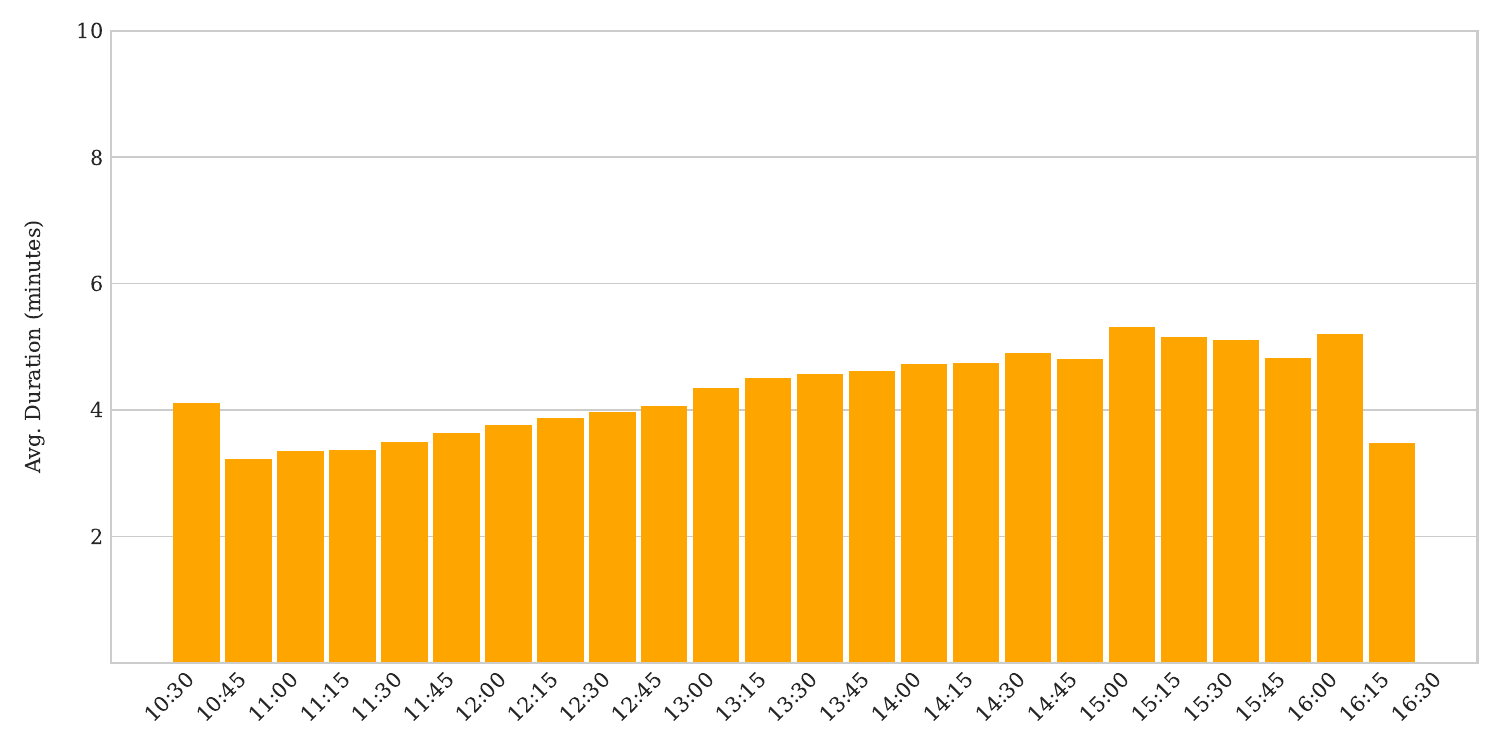}
  \caption{Mean duration of trades initiated during specific time intervals}
  \label{fig:intraday_duration}
\end{figure}

To further investigate the agent's intraday trading activity, we examine the distribution of the trades initiated within specific time intervals throughout the day. By aggregating all trades made by the agent across all assets, we illustrate the percentage of trades initiated within each 15-minute interval. This is visualized in Fig. \ref{fig:intraday_trades}. The agent initiates the majority of trades early in the day, gradually reducing its trading activity as the day progresses toward its conclusion. We can assume that the agent has learned that initiating trades later in the day entails higher levels of volatility.
Fig. \ref{fig:intraday_duration} visualizes the mean duration of trades initiated within specific time intervals throughout the day. The agent tends to hold positions for longer durations as the day progresses but reduces the trade duration in the final 15 minutes. This pattern indicates that the agent recognizes the significance of exiting positions before the market closes.

To validate the significance of positional features within our model, we conduct training on two model variants: one incorporating positional features and the other without. We also investigate the performance of our agent under different commissions used in the training and testing phase.  Table \ref{tab:without_pos} presents Sharpe values for both models across different commission fees. Our findings indicate that, on average, the model with positional features outperforms the alternative across higher commission fees. The model without positional features performs better only when no commission fees are assumed. Moreover, as commission increases, the model tends to become unprofitable, particularly after surpassing a $0.16$BP rate.
\begin{table}[]
\caption{Sharpe ratio of strategies with and without positional features across different commission fees}\label{tab:without_pos}
\begin{tabular}{lrrrrrr}
\toprule
   & \multicolumn{2}{c}{\textbf{COM=0BP}}                                            & \multicolumn{2}{c}{\textbf{COM=0.08BP}}                                         & \multicolumn{2}{c}{\textbf{COM=0.16BP}}                                         \\
   \midrule
   & \textbf{DRL ($\bm{\rho}_t$)} & \textbf{DRL ($\neg\bm{\rho}_t$)} & \textbf{DRL ($\bm{\rho}_t$)} & \textbf{DRL ($\neg\bm{\rho}_t$)} & \textbf{DRL ($\bm{\rho}_t$)} & \textbf{DRL ($\neg\bm{\rho}_t$)} \\
   \midrule
GC & 2.518 & \textbf{3.191} & \textbf{0.272} & 0.082 & \textbf{0.286} & -0.292\\
HG & 3.959 & \textbf{4.892} & \textbf{1.479} & 1.323 & \textbf{0.947} & 0.708 \\
SI & 3.109 & \textbf{3.326} & 1.293 & \textbf{1.308} & \textbf{0.764} & 0.507 \\
PL & 5.863 & \textbf{5.888} & \textbf{3.812} & 3.750 & 2.652 & \textbf{2.985} \\
CL & 1.117 & \textbf{1.341} & \textbf{0.791} & 0.429 & \textbf{0.607} & 0.309 \\
NG & 3.062 & \textbf{3.080} & \textbf{1.068} & 0.817 & \textbf{0.648} & 0.272 \\
\hline
A6 & 4.032 & \textbf{4.451} & \textbf{0.249} & 0.089 & \textbf{-0.470} & -1.225 \\
B6 & 3.340 & \textbf{3.658} & \textbf{0.048} & -0.07 & \textbf{-0.742} & -0.361 \\
E1 & 2.859 & \textbf{3.487} & \textbf{0.545} & -0.795 & \textbf{0.085} & -1.729 \\
MP & 3.113 & \textbf{3.223} & \textbf{0.327} & 0.151 & \textbf{0.586} & 0.217 \\
\bottomrule
\end{tabular}
\end{table}

\begin{figure}[H]
  \centering
  \includegraphics[height=0.3\textheight, width=1\textwidth]{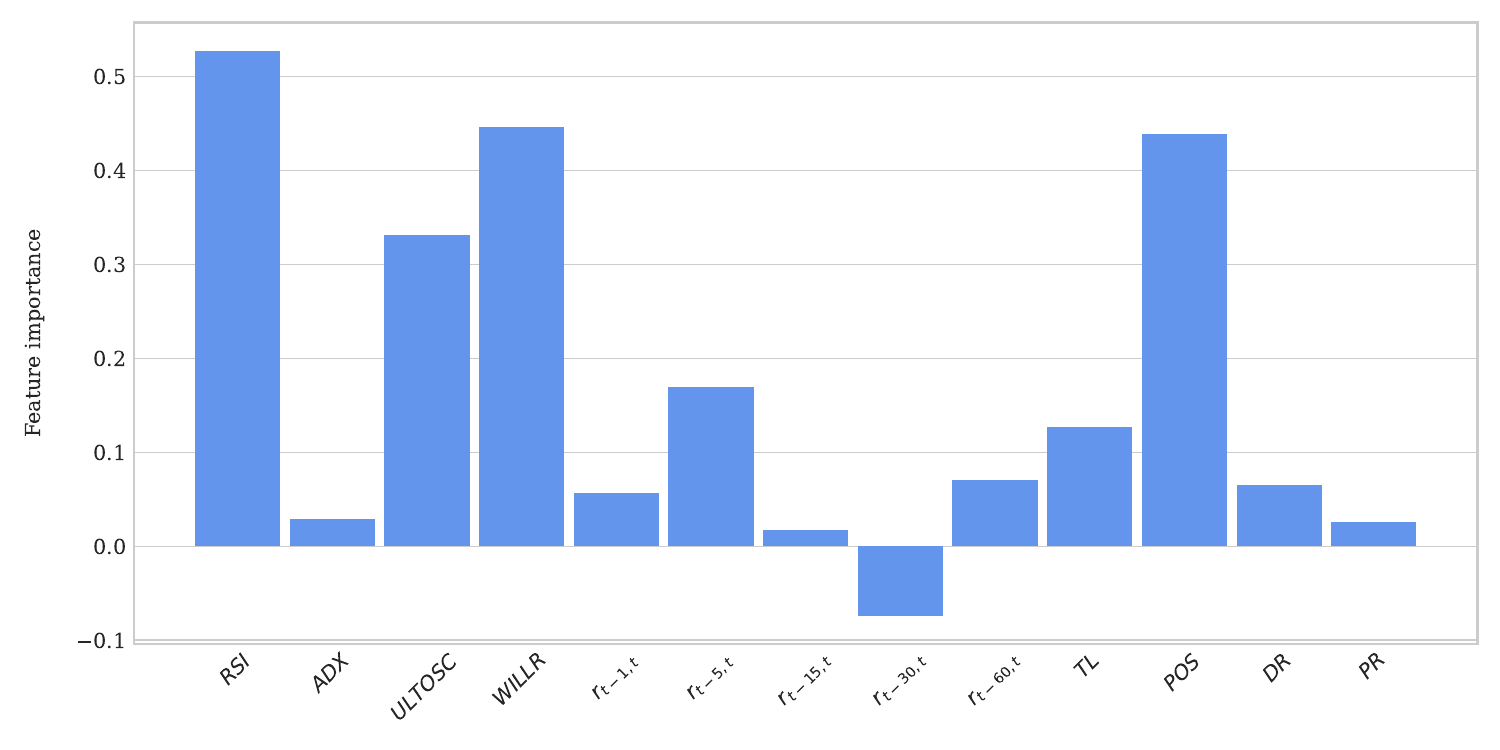}
  \caption{Individual feature contribution to the Sharpe ratio}
  \label{fig:feature_importance}
\end{figure}

Understanding which features significantly impact the agent's actions is crucial for gaining insights into his behavior. To investigate that, we explore the relative importance of each feature in influencing the agent's decision-making process \citep{Lim2019}. For each feature, we iterate over all assets and calculate the Sharpe ratio of the strategy with the corresponding feature set to zero throughout the agent's testing period. We compare this value to the Sharpe ratio of the agent's strategy where the agent observes all features, contained in Table \ref{tab:sharpe_assets}. We report feature importance as the mean difference between these two values. This method enables us to quantify feature importance based on the magnitude of performance reduction. Fig. \ref{fig:feature_importance} visually represents this analysis. All features play a role in the decision-making process, as there is a noticeable reduction in performance across all features. The only feature that appears to not improve the model is the half-hour ($r_{t-30,t}$) return feature. This analysis aligns with our decision to include positional features, highlighting their contribution to modeling the agent.

\section{Conclusion}\label{sec:conclusion}
This research paper presents a novel deep reinforcement learning (DRL) model tailored for intraday trading strategies. The model incorporates a sparsely structured state space enhanced with positional context, considering the agent's position relative to specific points in time. It is evaluated across various assets, including commodities and foreign exchange securities, over an extensive evaluation period of nearly a decade. Our results demonstrate the model's notable performance in terms of profitability and risk-adjusted metrics compared to established benchmarks. By quantifying the relative importance of each feature, we highlight the significant contribution of positional features to the agent's decision-making process. Moreover, we provide insights into the intraday trading activity exhibited by the agent. Our work extends the exploration of DRL applications in intraday trading, emphasizing the model's adaptability and robust performance across diverse assets.

Moving forward, there are promising directions for further research in the realm of DRL for intraday trading. Distributional reinforcement learning techniques could be explored to gain a more nuanced understanding of the entire distribution of returns, while hierarchical reinforcement learning models may offer a framework to capture multi-frequency decision-making processes. Model interpretability remains a critical research area as models are gaining acceptance in real-world applications. These avenues of research have the potential to enhance the robustness and generalization capabilities of DRL models for intraday trading.

\section*{Declarations}

\bmhead{Funding} This work was supported in part by the Croatian Science Foundation under Project 5241, and European Regional Development Fund under Grant KK.01.1.1.01.0009 (DATACROSS).

\bmhead{Data} Data is provided by a private data vendor at \href{https://firstratedata.com}{https://firstratedata.com}.

\bmhead{Competing interests} The authors have no competing interests to declare that are relevant to the content of this article.

\bmhead{Article Version} The Version of Record of this article is published in Evolving Systems, and is available online at \href{https://doi.org/10.1007/s12530-024-09593-6}{https://doi.org/10.1007/s12530-024-09593-6}

\bibliography{sn-bibliography}

\begin{thebibliography}{31}
\providecommand{\natexlab}[1]{#1}
\providecommand{\url}[1]{{#1}}
\providecommand{\urlprefix}{URL }
\providecommand{\doi}[1]{\url{https://doi.org/#1}}
\providecommand{\eprint}[2][]{\url{#2}}
 \bibcommenthead

\bibitem[{Bailey et~al(2014)Bailey, Borwein, de~Prado, and Zhu}]{Bailey2014}
Bailey DH, Borwein JM, de~Prado ML, et~al (2014) Pseudomathematics and financial charlatanism: The effects of backtest over fitting on out-of-sample performance. Notices of the AMS 61(5):458--471. \doi{10.1090/noti1105}

\bibitem[{Chan Phooi~M’ng and Mehralizadeh(2016)}]{Chan2016}
Chan Phooi~M’ng J, Mehralizadeh M (2016) Forecasting east asian indices futures via a novel hybrid of wavelet-pca denoising and artificial neural network models. PloS One 11(6):e0156338. \doi{10.1371/journal.pone.0156338}

\bibitem[{Cont(2011)}]{Cont2011}
Cont R (2011) Statistical modeling of high-frequency financial data. IEEE Signal Processing Magazine 28(5):16--25. \doi{10.1109/MSP.2011.941548}

\bibitem[{Deng et~al(2016)Deng, Bao, Kong, Ren, and Dai}]{Deng2016}
Deng Y, Bao F, Kong Y, et~al (2016) Deep direct reinforcement learning for financial signal representation and trading. IEEE Transactions on Neural Networks and Learning Systems 28(3):653--664. \doi{10.1109/TNNLS.2016.2522401}

\bibitem[{Golu{\v{z}}a et~al(2023)Golu{\v{z}}a, Bauman, Kova{\v{c}}evi{\'c}, and Kostanj{\v{c}}ar}]{Goluza2023}
Golu{\v{z}}a S, Bauman T, Kova{\v{c}}evi{\'c} T, et~al (2023) Imitation learning for financial applications. In: 2023 46th MIPRO ICT and Electronics Convention (MIPRO), pp 1130--1135

\bibitem[{Gurrib et~al(2018)}]{Gurrib2018}
Gurrib I, et~al (2018) Performance of the average directional index as a market timing tool for the most actively traded usd based currency pairs. Banks and Bank Systems 13(3):58--70. \doi{10.21511/bbs.13(3).2018.06}

\bibitem[{Hendershott et~al(2011)Hendershott, Jones, and Menkveld}]{Hendershott2011}
Hendershott T, Jones CM, Menkveld AJ (2011) Does algorithmic trading improve liquidity? The Journal of Finance 66(1):1--33. \doi{10.1111/j.1540-6261.2010.01624.x}

\bibitem[{Huang(2018)}]{Huang2018}
Huang CY (2018) Financial trading as a game: A deep reinforcement learning approach Preprint at \url{https://arxiv.org/abs/1807.02787}

\bibitem[{Huddleston et~al(2023)Huddleston, Liu, and Stentoft}]{Huddleston2023}
Huddleston D, Liu F, Stentoft L (2023) Intraday market predictability: A machine learning approach. Journal of Financial Econometrics 21(2):485--527. \doi{10.1093/jjfinec/nbab007}

\bibitem[{Kingma and Ba(2014)}]{Kingma2014}
Kingma DP, Ba J (2014) Adam: A method for stochastic optimization Preprint at \url{https://arxiv.org/abs/1412.6980}

\bibitem[{Kolm et~al(2023)Kolm, Turiel, and Westray}]{Kolm2023}
Kolm PN, Turiel J, Westray N (2023) Deep order flow imbalance: Extracting alpha at multiple horizons from the limit order book. Mathematical Finance 33(4):1044--1081. \doi{10.1111/mafi.12413}

\bibitem[{Kova{\v{c}}evi{\'c} et~al(2022)Kova{\v{c}}evi{\'c}, Golu{\v{z}}a, Mer{\'c}ep, and Kostanj{\v{c}}ar}]{Kovavcevic2022}
Kova{\v{c}}evi{\'c} T, Golu{\v{z}}a S, Mer{\'c}ep A, et~al (2022) Effect of labeling algorithms on financial performance metrics. In: 2022 45th Jubilee International Convention on Information, Communication and Electronic Technology (MIPRO), pp 980--984

\bibitem[{Li et~al(2023)Li, Khishe, and Qian}]{Li2023}
Li X, Khishe M, Qian L (2023) Evolving deep gated recurrent unit using improved marine predator algorithm for profit prediction based on financial accounting information system. Complex \& Intelligent Systems 10:595--611. \doi{https://doi.org/10.1007/s40747-023-01183-4}

\bibitem[{Li et~al(2019)Li, Zheng, and Zheng}]{Li2019}
Li Y, Zheng W, Zheng Z (2019) Deep robust reinforcement learning for practical algorithmic trading. IEEE Access 7:108014--108022. \doi{10.1109/ACCESS.2019.2932789}

\bibitem[{Lim et~al(2019)Lim, Zohren, and Roberts}]{Lim2019}
Lim B, Zohren S, Roberts S (2019) Enhancing time-series momentum strategies using deep neural networks. The Journal of Financial Data Science \doi{10.3905/jfds.2019.1.015}

\bibitem[{Liu et~al(2021)Liu, Li, Li, Li, and Xie}]{Liu2021}
Liu F, Li Y, Li B, et~al (2021) Bitcoin transaction strategy construction based on deep reinforcement learning. Applied Soft Computing 113:107952. \doi{https://doi.org/10.1016/j.asoc.2021.107952}

\bibitem[{Liu et~al(2020)Liu, Liu, Zhao, Pan, and Liu}]{Liu2020}
Liu Y, Liu Q, Zhao H, et~al (2020) Adaptive quantitative trading: An imitative deep reinforcement learning approach. In: Proceedings of the AAAI conference on artificial intelligence, pp 2128--2135

\bibitem[{Moody and Saffell(2001)}]{Moody2001}
Moody J, Saffell M (2001) Learning to trade via direct reinforcement. IEEE Transactions on Neural Networks 12(4):875--889. \doi{10.1109/72.935097}

\bibitem[{Moskowitz et~al(2012)Moskowitz, Ooi, and Pedersen}]{Moskowitz2012}
Moskowitz TJ, Ooi YH, Pedersen LH (2012) Time series momentum. Journal of Financial Economics 104(2):228--250. \doi{10.1016/j.jfineco.2011.11.003}

\bibitem[{Murphy(1999)}]{Murphy1999}
Murphy JJ (1999) Technical analysis of the financial markets: A comprehensive guide to trading methods and applications. New York Institute of Finance, New York, NY, USA

\bibitem[{Narang(2013)}]{Narang2013}
Narang RK (2013) Inside the black box: A simple guide to quantitative and high frequency trading, vol 846. John Wiley \& Sons, Hoboken, NJ, USA

\bibitem[{Poterba and Summers(1988)}]{Poterba1988}
Poterba JM, Summers LH (1988) Mean reversion in stock prices: Evidence and implications. Journal of Financial Economics 22(1):27--59. \doi{10.1016/0304-405X(88)90021-9}

\bibitem[{Schulman et~al(2015)Schulman, Moritz, Levine, Jordan, and Abbeel}]{Schulman2015}
Schulman J, Moritz P, Levine S, et~al (2015) High-dimensional continuous control using generalized advantage estimation Preprint at \url{https://arxiv.org/abs/1506.02438}

\bibitem[{Schulman et~al(2017)Schulman, Wolski, Dhariwal, Radford, and Klimov}]{Schulman2017}
Schulman J, Wolski F, Dhariwal P, et~al (2017) Proximal policy optimization algorithms Preprint at \url{https://arxiv.org/abs/1707.06347}

\bibitem[{Si et~al(2017)Si, Li, Ding, and Rao}]{Si2017}
Si W, Li J, Ding P, et~al (2017) A multi-objective deep reinforcement learning approach for stock index future’s intraday trading. In: 2017 10th International symposium on computational intelligence and design (ISCID), pp 431--436

\bibitem[{Sun et~al(2022)Sun, Xue, Wang, He, Zhu, Li, and An}]{Sun2022}
Sun S, Xue W, Wang R, et~al (2022) Deepscalper: A risk-aware reinforcement learning framework to capture fleeting intraday trading opportunities. In: Proceedings of the 31st ACM International Conference on Information \& Knowledge Management, pp 1858--1867

\bibitem[{Sutton and Barto(2018)}]{Sutton2018}
Sutton RS, Barto AG (2018) Reinforcement learning: An introduction. MIT press, Cambridge, MA, USA

\bibitem[{Vojtko and Padysak(2020)}]{Vojtko2020}
Vojtko R, Padysak M (2020) Continuous futures contracts methodology for backtesting Preprint at \url{https://ssrn.com/abstract=3517736}

\bibitem[{Wen et~al(2022)Wen, Bouri, Xu, and Zhao}]{Wen2022}
Wen Z, Bouri E, Xu Y, et~al (2022) Intraday return predictability in the cryptocurrency markets: Momentum, reversal, or both. The North American Journal of Economics and Finance 62:101733. \doi{10.1016/j.najef.2022.101733}

\bibitem[{Yang et~al(2020)Yang, Liu, Zhong, and Walid}]{Yang2020}
Yang H, Liu XY, Zhong S, et~al (2020) Deep reinforcement learning for automated stock trading: An ensemble strategy. In: Proceedings of the first ACM international conference on AI in finance, pp 1--8

\bibitem[{Zhang et~al(2020)Zhang, Zohren, and Stephen}]{Zhang2020}
Zhang Z, Zohren S, Stephen R (2020) Deep reinforcement learning for trading. The Journal of Financial Data Science \doi{10.3905/jfds.2020.1.030}

\end{thebibliography}

\end{document}